\documentclass[a4paper,11pt]{article}
\pdfoutput=1 

\usepackage{jinstpub} 

\title{Investigation of neutron scattering in the Multi-Blade detector with Geant4 simulations}
\usepackage[normalem]{ulem}

\usepackage{placeins} 
\usepackage{subcaption} 
\usepackage{booktabs,makecell, diagbox} 
\usepackage[table]{xcolor} 

\graphicspath{{plots/}}
\DeclareGraphicsExtensions{.pdf,.jpeg,.png,.jpg}

\author[a,b]{G. Galg\'oczi}
\author[c,1]{K. Kanaki%
\note{Corresponding author.}}
\author[c]{F. Piscitelli}
\author[c]{T. Kittelmann}
\author[b]{D. Varga}
\author[c,d]{R. Hall-Wilton}

\affiliation[a]{E\"otv\"os Lor\'and University, 1053 Budapest, Egyetem
  t\'er 1-3., Hungary}
\affiliation[b]{Hungarian Academy of Sciences, Wigner Research Centre for Physics, 1525 Budapest 114., Hungary}
\affiliation[c]{European Spallation Source ESS ERIC, SE-221 00 Lund, Sweden}
\affiliation[d]{Mid-Sweden University, SE-851 70 Sundsvall, Sweden}

\emailAdd{Kalliopi.Kanaki@esss.se}

\abstract{
The European Spallation Source (ESS) is the world's next generation
spallation-based neutron source. The research conducted at ESS will
yield in the discovery and development of new materials including the
fields of manufacturing, pharmaceuticals, aerospace, engines,
plastics, energy, telecommunications, transportation, information
technology and biotechnology. The spallation source will deliver an
unprecedented neutron flux. In particular, the reflectometers selected
for construction, ESTIA and FREIA, have to fulfill challenging requirements. Local
incident peak rate can reach 10$^5$~Hz/mm$^2$. For new science to be
addressed, the spatial resolution is aimed
to be less than 1~mm with a desired scattering of 10$^{-4}$ (peak-to-tail ratio). The
latter requirement is approximately two orders of magnitude better than the
current state-of-the-art detectors. The main aim of this work is to
quantify the cumulative contribution of various detector components to the scattering
of neutrons and to prove that the respective effect is within
the requirements set for the Multi-Blade detector by the ESS
reflectometers. To this end, different sets of geometry and beam parameters are investigated,
with primary focus on the cathode coating and the detector window thickness.

}


\begin{document}
\maketitle
\flushbottom

\section{Introduction}

The ESS ERIC~\cite{ess}, currently under
construction in Lund, Sweden, aspires to become the most powerful
pulsed neutron source in the world. With its long pulse of 2.86~ms and
a brilliance higher than 10$^{14}$~n/cm$^2$/s/sr/\AA, it can deliver unprecedented flux on
the sample and revolutionise the way neutron experiments are conducted~\cite{esstdr,garoby}. The produced neutrons are destined to serve a variety of instruments for
reflectometry, diffraction, spectrometry and imaging
purposes. According to the current schedule the first neutrons will hit the target in 2019, with the user programme starting in 2023. 

Reflectometry is an experimental technique present at every
neutron source. Hence, two of the first instruments approved for construction at ESS are
reflectometers. The one, with a vertical scattering plane, is called the
Fast Reflectometer for Extended Interfacial Analysis
(FREIA)~\cite{freia,freia2,freia3} and the horizontal one is ESTIA~\cite{estia,estia2,estia3}. The
former one will be optimised for magnetic samples and in-situ or
in-operando studies. The latter reflectometer is designed to achieve
the best performance for liquid/liquid or liquid/gas interfaces. With
a sample flux of 10$^9$-10$^{10}$~n/s/cm$^2$ and a high sample
reflectivity ($\sim$90\%) it is expected that the peak instantaneous
rate of the neutron detectors could reach
100~kHz/mm$^{2}$~\cite{estia,estia2,estia3}. This value exceeds the rate capability
of current neutron detector technologies (including $^{3}$He-based
detectors) by approximately two orders of magnitude. Additionally, the limit set for neutron scattering inside the detector is lower than what the current state-of-the-art detectors are capable of~\cite{estia_prop}. These are the two biggest challenges for the detector design aimed at reflectometry for ESS.

In the past $^{3}$He-based neutron detectors played a key role for
thermal and cold neutron detection~\cite{illbb}. Due to the limitations of these detectors in
scientific performance and the shortage of~$^{3}$He~\cite{he_shortage,he_shortage2}, the focus of the neutron
detector community has shifted to alternative, higher-performing solutions, such as
$^{10}$B$_4$C-based
detectors~\cite{kirstein2014,multiwire_2013,lacy2013,mgcncs,bandgem4,rpc,rpc_arxiv},
scintillators~\cite{isis1,isis2,wang11,nop_sonde,sonde_arxiv,SCI_katagari,jparc,SCI_bell}
or$^6$LiF-based solid state silicon detectors~\cite{lif,ear1,sintef}. Additionally, it is already proven that $^{10}$B$_4$C-based detectors are capable of outperforming $^{3}$He detectors in terms of spurious scattering of neutrons~\cite{estia_prop, MB2017}. Several studies demonstrate the performance of this detector type and its suitability for neutron scattering experiments~\cite{MB2017,crisp,mauri2,phd_piscitelli}.

The detector design developed and adopted for the ESS reflectometers
is the Multi-Blade detector~\cite{MB2017,crisp,mauri2,phd_piscitelli,royalsoc,MBconceptFP}. It has been extensively characterised
and validated at various neutron facilities and with various types of
samples. To get a deeper understanding of the scattering
patterns the detector geometry causes and to support the choice of
materials and component dimensions, a detailed Geant4~\cite{geant4a,geant4b,geant4c} detector model is implemented. The
main aim of this simulation effort is to prove that the Multi-Blade detector
meets the requirements set by ESS, particularly the scattering of
$10^{-4}$~(see p.9, figure 9.1 in~\citep{estia2}). 

In the following sections the Multi-Blade detector is introduced. The
details of the Geant4 implementation are presented, as well as the
figure of merit used in the subsequent analysis. Fractional
scattering is defined and sources of background caused by misplaced
detection events are identified and studied as a function of neutron
wavelength. The results are discussed with respect to the instrument
requirements.

\FloatBarrier

\section{The Multi-Blade model in Geant4}

\subsection{The Multi-Blade detector}
\label{sec:mb}

\begin{figure}[!h]   
  \centering
  \begin{subfigure}{0.445\textwidth}
    \centering
    \includegraphics[width=\textwidth, angle=270]{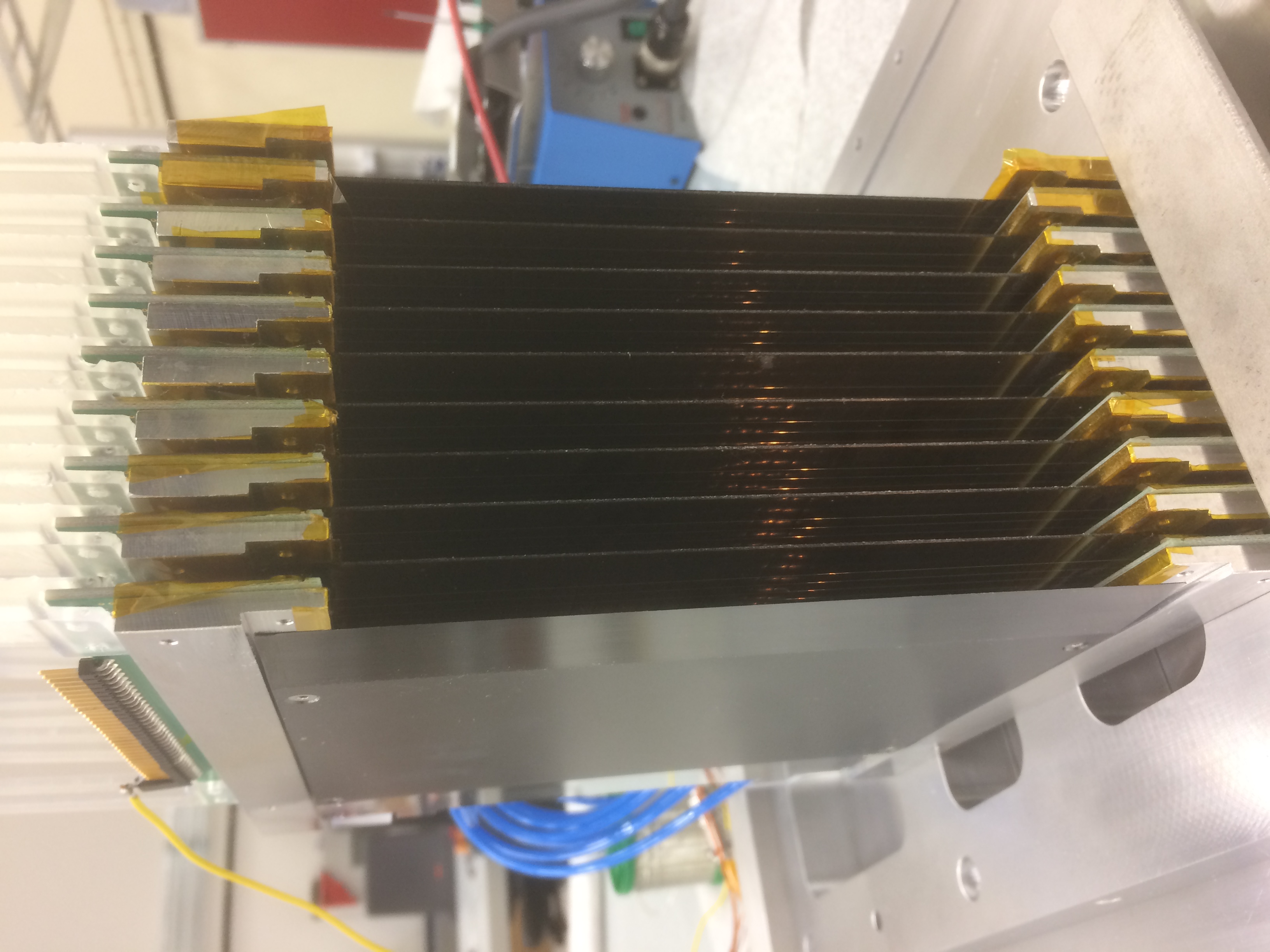}
    \caption{\footnotesize}
    \label{mbPhoto}    
  \end{subfigure}%
  \begin{subfigure}{0.555\textwidth}
    \centering
    \includegraphics[width=\textwidth]{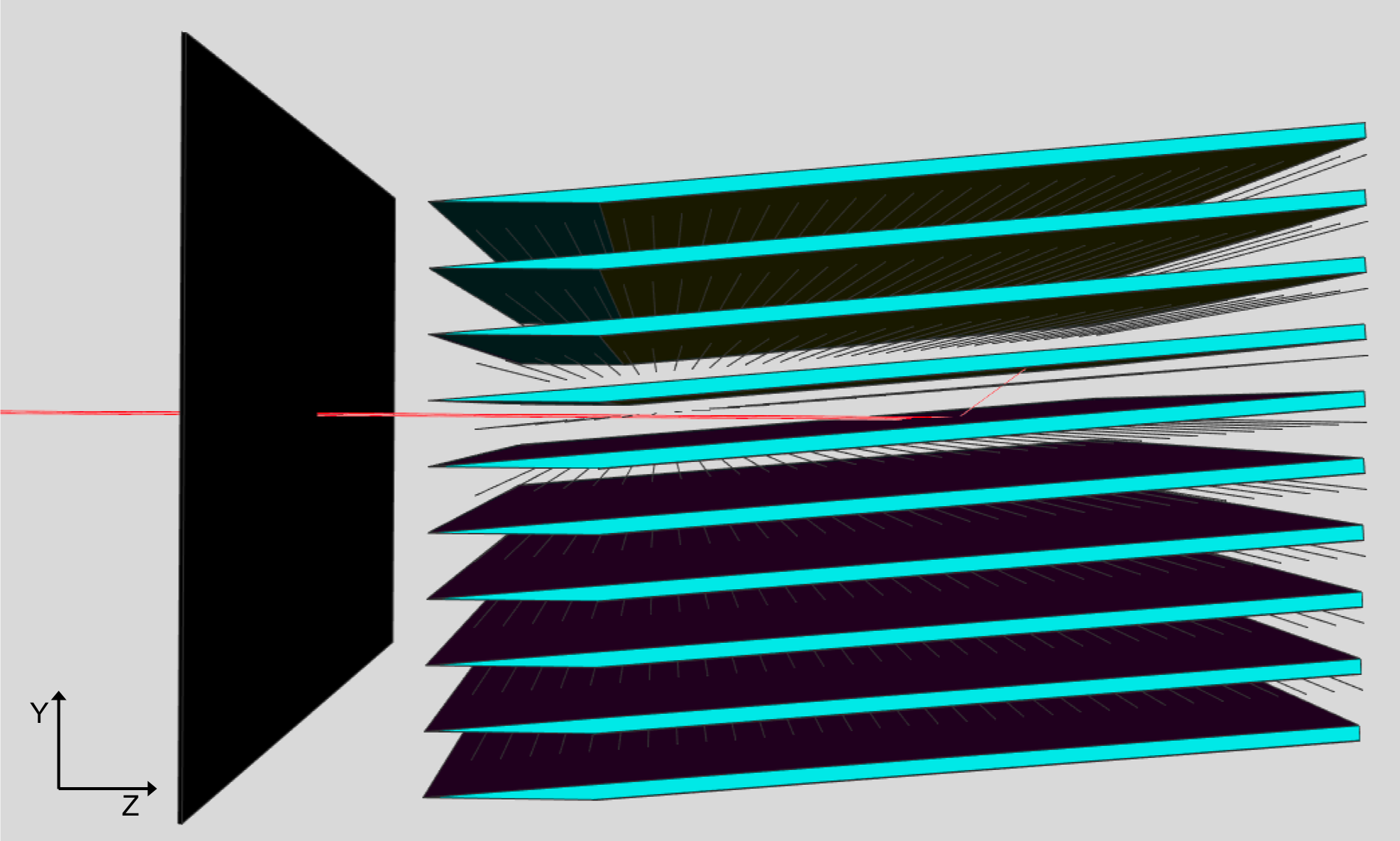}
    \caption{\footnotesize}
    \label{mbModel}
  \end{subfigure}
  \caption{\footnotesize (a) A prototype of the Multi-Blade
    detector~\cite{crisp}. (b) Geant4 geometry model being hit by a red neutron
    beam through the detector entrance window. The vessel is
    absent from both figures to allow a clear view of the geometry.}
  \label{}
\end{figure}

The Multi-Blade detector is a novel neutron detector currently being
designed at ESS (see figure~\ref{mbPhoto}). Its development was initiated at Institute
Laue-Langevin (ILL)~\citep{ill,MBconceptFP,MBconcept,phd_piscitelli}. It consists of a set of successive
Multi-Wire Proportional Chambers~\cite{mwpc}, with
$^{10}$B$_4$C-coated cathodes and Ar/CO$_2$ (80/20 by volume) as a gas counter
mixture. In addition to the anode wires, each chamber is equipped with
a strip readout. The strips are perpendicular to the wire direction,
in order to achieve two-dimensional spatial resolution. 

The detector consists of cassettes as building units. Each one has a
boron carbide ($^{10}$B$_4$C) converter layer that is on a titanium
substrate called a ``blade''. On the other side of the blade a kapton
layer and 32 copper strips are located for charge readout. Between the
blades 32 wires are stretched at a 4~mm pitch. The blades are tilted in a way that the
incoming neutron beam hits the converter layer with an incident angle
of 5$^\circ$, as shown in figure~\ref{mbModel}. Therefore the
thickness of the converter is viewed by the neutrons as being
effectively about 11 times larger~\cite{Absorbing_Film,MB2017}. Several cassettes are assembled together forming a fan-like arrangement to achieve the area coverage required for reflectometry.

\subsection{Implementation of the detector model in Geant4} \label{sec:sim}


It is exactly this fan-like geometry that is the core of the current
simulation study. The detector geometry and data analysis are
implemented in a framework based on Geant4 developed by the ESS Detector
Group~\cite{ess_coding_framework,icns}. The detector geometry consists of
ten 130~mm~$\times$~140~mm~$\times$~2~mm titanium blades coated with enriched $^{10}$B$_4$C by 98\%
on one side and a kapton and copper layer on the other, with
thicknesses of 30~$\mu$m and 40~$\mu$m respectively. The copper layer
is not segmented in strips, unlike the real prototype, as the distance between the strips is 0.1-0.2~mm and the effect of spurious scattering in this part of the detector is negligible compared to other materials. The tungsten wires are also included in
the model. The Ar/CO$_2$ mixture has a ratio of 80/20 by
volume and a pressure of 1.1~bar. The gas vessel surrounds the entire
detector structure. It consists of enriched $^{10}$B$_4$C acting as a
total absorber, in order to prevent
scattered neutrons from entering the active volume again. On the
beam entrance side an aluminium window allows the neutrons to reach
the converter (see figure~\ref{mbModel}).

All materials are selected from the Geant4 database of NIST materials, except for Ti, Al, Cu and W. The latter are
described with the use of the NCrystal library~\cite{ncrystal,icns}, as their crystalline structure is important for the correct treatment of their interaction with thermal neutrons. The physics list
used is QGSP\_BIC\_HP and the Geant4 version is 10.00.p03.

Last, but not least, the neutron generator is a mono-energetic pencil beam, impinging the
converter layer at a 5$^{\circ}$ angle in the centre of the middle
blade, as in figure~\ref{mbModel}. The generator parameters, albeit
simple, condense the characteristics of typical neutron
distributions from reflectometry samples that are important for this
study. All results are produced with 1 million events.

\subsection{Implementation of the detection process in the simulation}
\label{sec:hit}

In boron-based neutron detectors the secondary charged particles
($\alpha$ and $^7$Li ions) are created after the $^{10}$B nucleus
captures a neutron. The cross section of this process is 3835~b for
neutrons with a wavelength of 1.8~\AA~(p.~15, \cite{illbb}). The position where the neutron is absorbed is
called the conversion point. Therefore the neutrons are detected
indirectly. The ion products cross the converter and enter the gas
volume with a remaining energy. The maximum distance the $^7$Li and $\alpha$ ions can
travel in the specific counting gas is 3~mm and 6~mm respectively at
atmospheric pressure and room temperature conditions.  

\begin{figure}[!h]
  \centering
  \includegraphics[width=.6\textwidth]{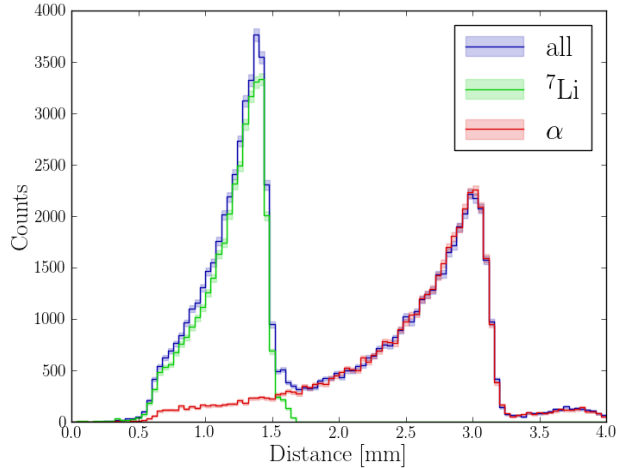}
  \caption{\footnotesize Distance between conversion and hit position
    as approximated in a simulation for 12~\AA~neutrons. The two peaks reflect the maximum path of $^7$Li and $\alpha$ ions
    respectively, but are halved in this representation (1.5~mm and
    3~mm) as a result of the hit definition. The bands around the lines represent the statistical
  uncertainties.} 
  \label{conv_hit_dist}    
\end{figure}

In order to describe the physical phenomena behind the detection of
neutrons in the Multi-Blade detector, an approximation is used. The
geometrical center of the tracks of the charged particles created by
ionisation in the gas of the detector is defined as the hit position
(see figure~\ref{conv_hit_dist}). Furthermore, the fact that the readout of the hits is done by wires and
strips has to be taken into account. For example, in the direction
perpendicular to the blades surface, the coordinate of the hit is measured by
the wires. Therefore the position of the triggered wire is read out,
not the actual position of the centre of the charge cloud. In order to
emulate this effect, the hits are projected to the plane of the
boron surface, a transformation that brings the simulation
treatment closer to the experimental process.

\FloatBarrier

\subsection{Comparison of detection efficiency obtained by measurements and Geant4}
\label{sec:validation}

The simulation results produced with the ESS simulation framework have
been previously validated in~\cite{mg_scattering}. An additional validation
of the current model is performed in terms of detection
efficiency. The results are compared to analytical calculations~\citep{MBconceptFP} and
experimental measurements performed at the ATHOS instrument of the BNC facility~\citep{MB2017}
in Hungary and CRISP~\cite{crisp,royalsoc} at ISIS, UK. 
\begin{figure}[!h]
  \centering   
  \includegraphics[width=.6\textwidth]{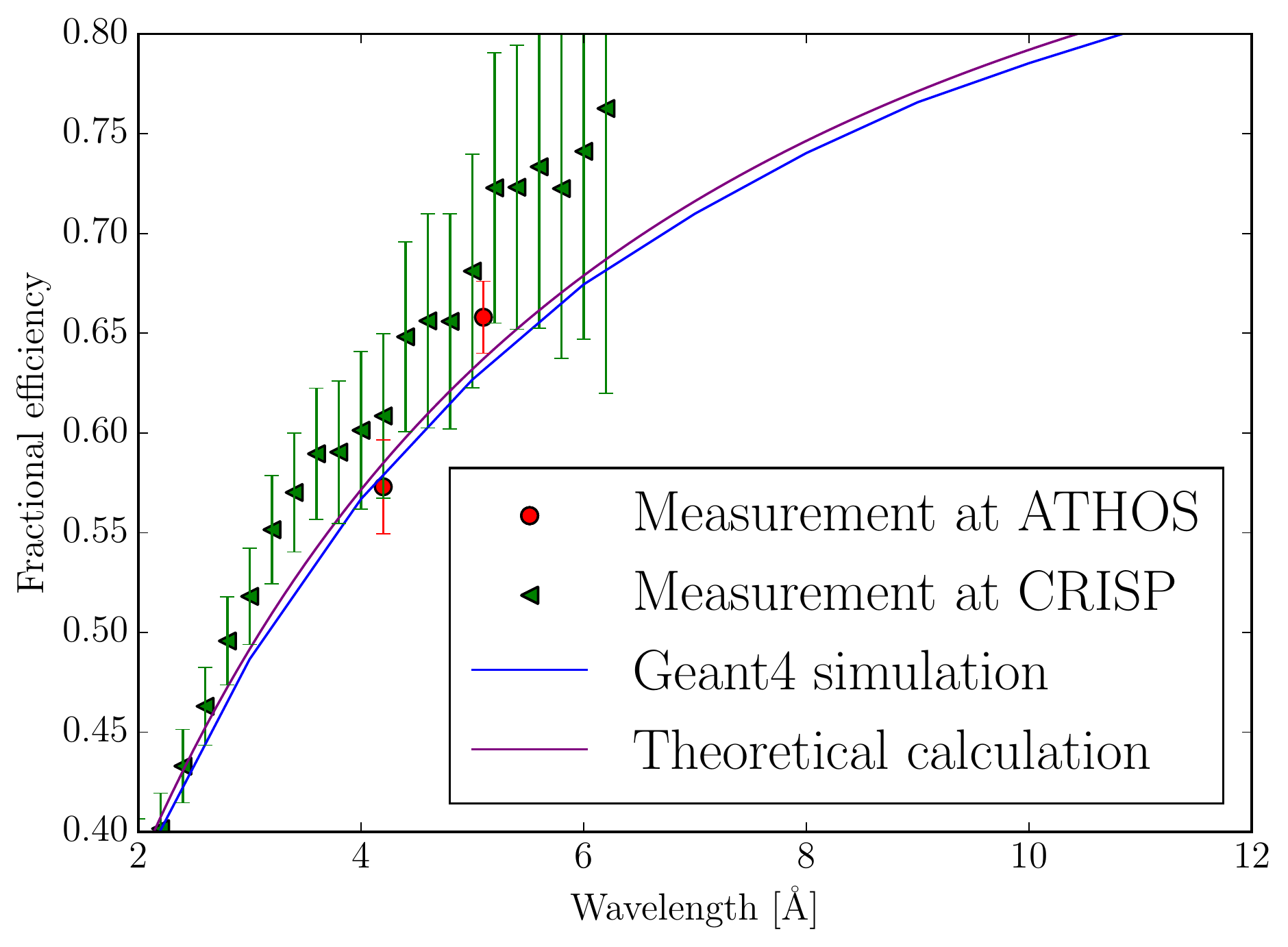}
  \caption{\footnotesize Fractional detection efficiency of the Multi-Blade detector as a function of
    neutron wavelength, obtained from theoretical
    calculations~\cite{MBconceptFP,alvaro}, measurements~\cite{MB2017,crisp}
    and Geant4 simulation (this work).}
  \label{eff}    
\end{figure}
The observed agreement is within the error bars for most of the
neutron wavelengths. The uncertainties of the experimental points depend on the statistics of each measurement and systematic effects on the
experimental data are not accounted for in the simulation. In
addition, the efficiency estimate of the simulation is affected to a
small extent by the fact that the detector gas volume is not segmented
in the simulation. The
detection threshold of 120~keV on the energy deposition is applied
per event for the entire gas volume and not per wire, as in the
experimental data. However, due to the localised nature of
the energy deposition the approximation does not compromise the
simulation for the purposes of this work. 

\FloatBarrier

\section{Scattering effects}  
\label{sec:results}

\subsection{Definition of spurious detection events}
\label{sec:misplaced}

\begin{figure}[!h]   
  \centering
  \begin{subfigure}{0.7\textwidth}
    \centering
    \includegraphics[width=\textwidth]{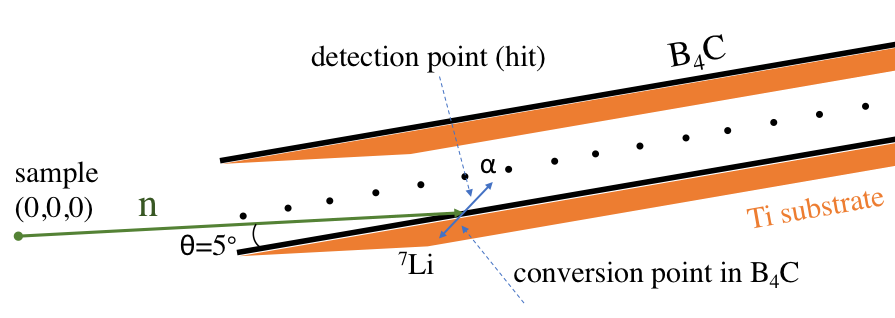}
    \caption{\footnotesize }
    \label{resolution}    
  \end{subfigure}
  \begin{subfigure}{0.5\textwidth}
    \centering
    \includegraphics[width=\textwidth]{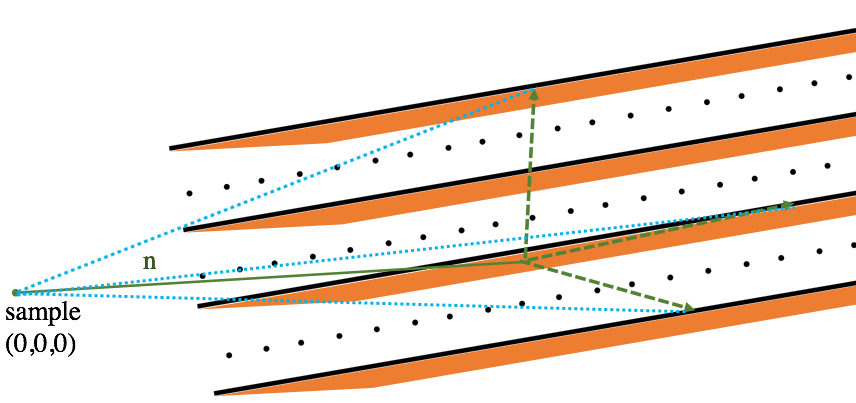}
    \caption{\footnotesize }
    \label{scatteringBlade}
  \end{subfigure}%
  \begin{subfigure}{0.5\textwidth}
    \centering
    \includegraphics[width=\textwidth]{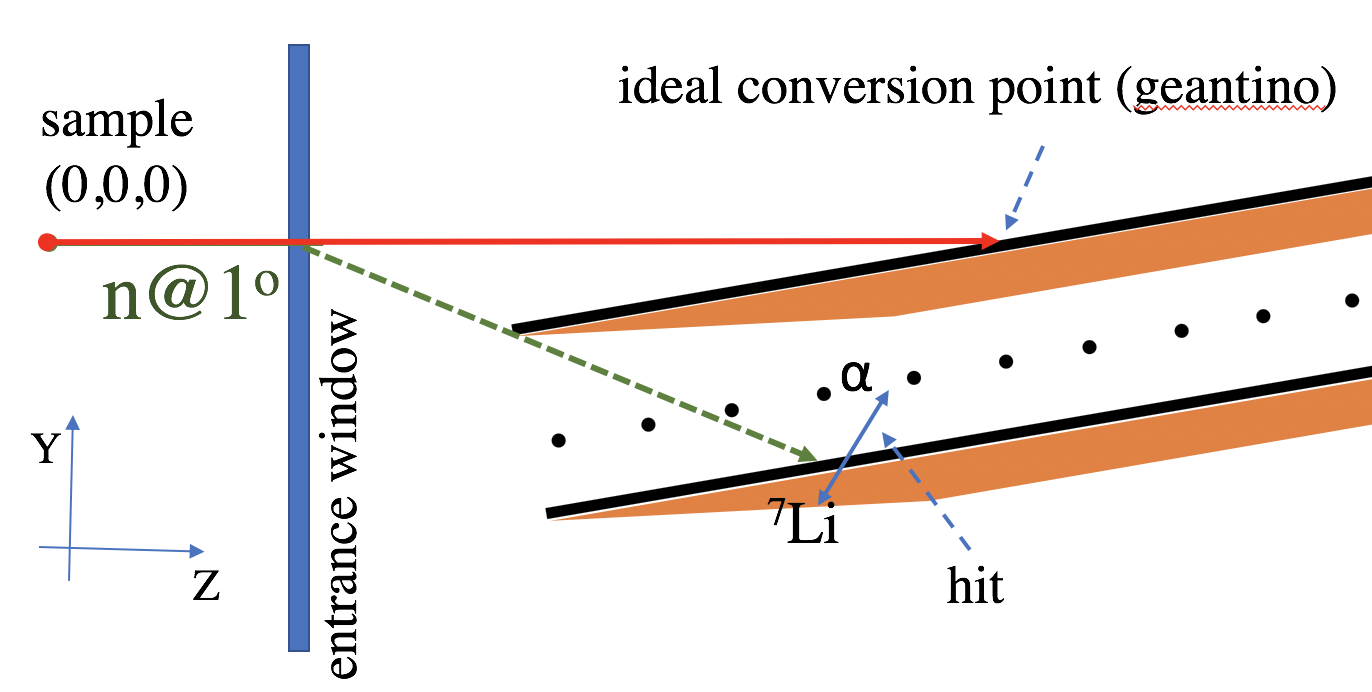}
    \caption{\footnotesize }
    \label{scatteringWindow}
  \end{subfigure}
  \caption{\footnotesize (a) Difference between conversion and detection
    point. (b) A neutron
    traversing the first converter layer (solid green) can scatter in the blade
    material and finally get converted away from the first crossing
    point (dashed green). This leads to the miscalculation of the distance between sample and
    detection point (dashed blue). (c) Similarly for a scattered neutron on the
    detector window. The latter is 1$^{\circ}$ inclined with respect to
  the vertical axis. The projection of the detection point on the
  converter layer is not displayed here for view
  simplification.}
  \label{scatteringExamples}
\end{figure}

The Multi-Blade is a position sensitive detector, intended to operate
in Time-of-Flight (TOF) mode. This means that the energy of the
incident neutron is indirectly derived from a time and a 3D position
measurement. Two factors can impact the precision of the neutron position reconstruction;
the detector spatial resolution (short-scale effect) and scattering
(long-scale effect), with different impact to the distribution of the
detection coordinates. Figure~\ref{scatteringExamples} demonstrates
these scenarios. In figure~\ref{resolution} a neutron,
which is absorbed in the first converter layer it meets, leads to a secondary
particle releasing its energy in the counting gas (the other particle
gets stopped inside the cathode substrate and is therefore lost). The spatial
resolution of the detector determined by the anode wire pitch and the
strip width locally smears the experimental detection point in the
data reconstruction process. Figures~\ref{scatteringBlade} and \ref{scatteringWindow} depict long-range
effects stemming from neutron scattering either within the detector itself
or in the entrance window respectively. Such events lead to the
miscalculation of the distance between sample and detection point, and
eventually to a wrongly derived value for the incident neutron energy
and scattering vector.

In order to quantify the impact of the misplaced detected neutrons, an
nonphysical technical Geant4 particle, called
``geantino'', is utilised. This particle is generated along with each
primary neutron with the same initial parameters (see figure~\ref{scatteringWindow}). Geantinoes do not
interact with matter, therefore their tracks are straight lines. The
point where a primary neutron would be detected in an ideal
measurement is defined by the intersection of the respective
geantino track with the first converter layer it traverses. This
position is then compared against the respective neutron detection
coordinates. The definition of this condition is motivated by the
detector design; the actual thickness of the converter is
at least 7.5~$\mu$m~\cite{MB2017,crisp} to ensure that, aside from a high detection efficiency, almost all neutrons convert in the first
boron layer they encounter, thus reducing the scattering caused by the
cathode material.

The utilisation of geantinoes allows for this study to be realised for arbitrary generators and geometries. However, in the
current one, due to the fact that primary neutrons start
from (0,0,0) and form a pencil beam on the Z axis, the geantino ``detection'' coordinates are also 0 in X
and Y. This means that the actual
neutron hit coordinates are sufficient for the visualisation of
scattering effects in this particular case ($\rm
X_{hit}-X_{geantino}=X_{hit}$, $\rm
Y_{hit}-Y_{geantino}=Y_{hit}$). These hit coordinates are the ones
projected on the converter layer as explained at the end of
subsection~\ref{sec:hit}.

The contributors to the scattering effects studied in this work are
the converter layer and the entrance window of the detector, both
looked at as a function of neutron wavelength. The material (Ti) and
thickness of the blade (2~mm) have been dictated by engineering
needs and the coating process and are fixed for all simulations of this work.

\begin{figure}[!h]   
  \centering
  \begin{subfigure}{0.6\textwidth}
    \centering
    \includegraphics[width=\textwidth]{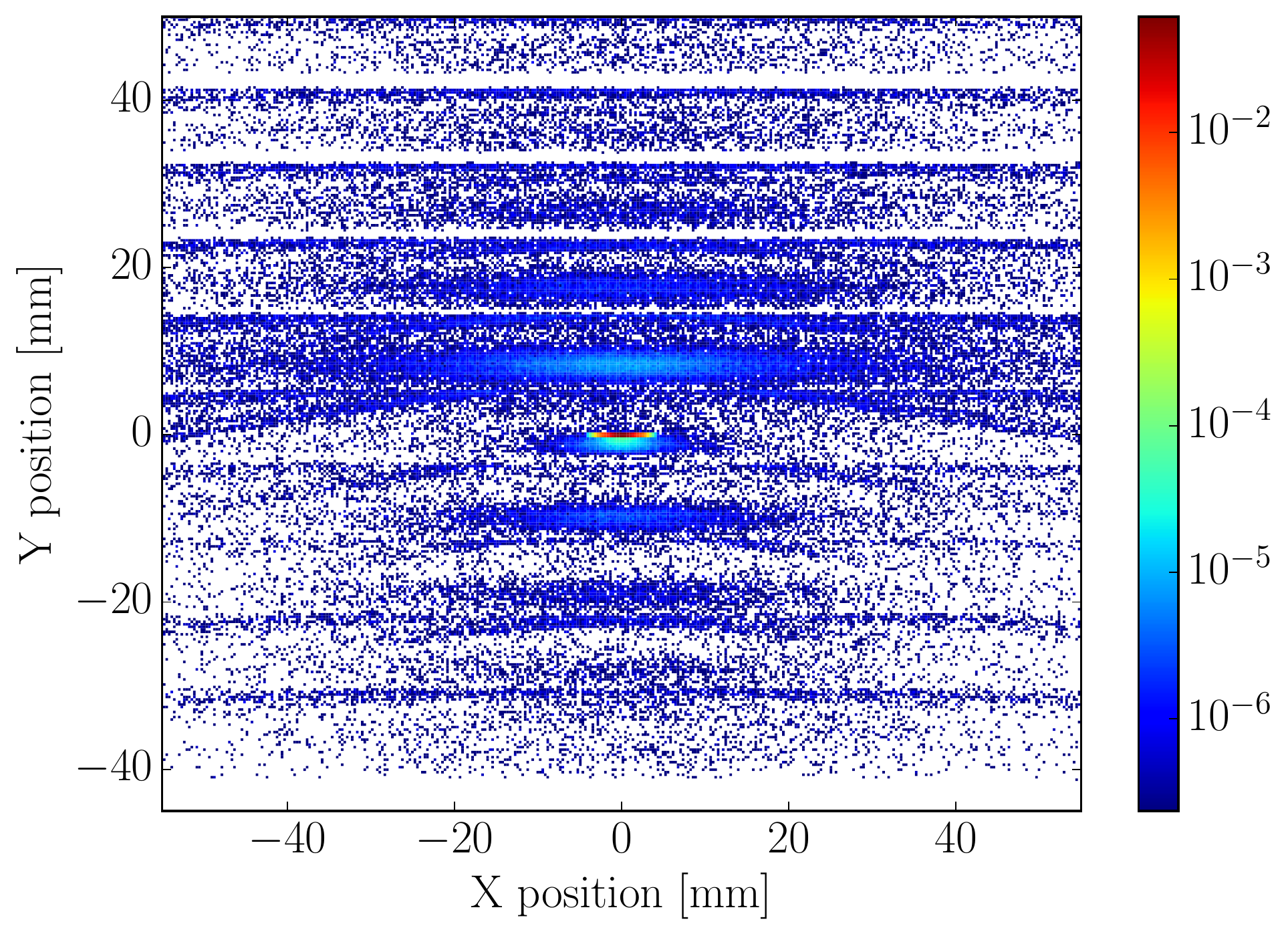}
    \caption{\footnotesize}
    \label{fig:2dhits}    
  \end{subfigure}
  \begin{subfigure}{0.5\textwidth}
    \centering
    \includegraphics[width=\textwidth]{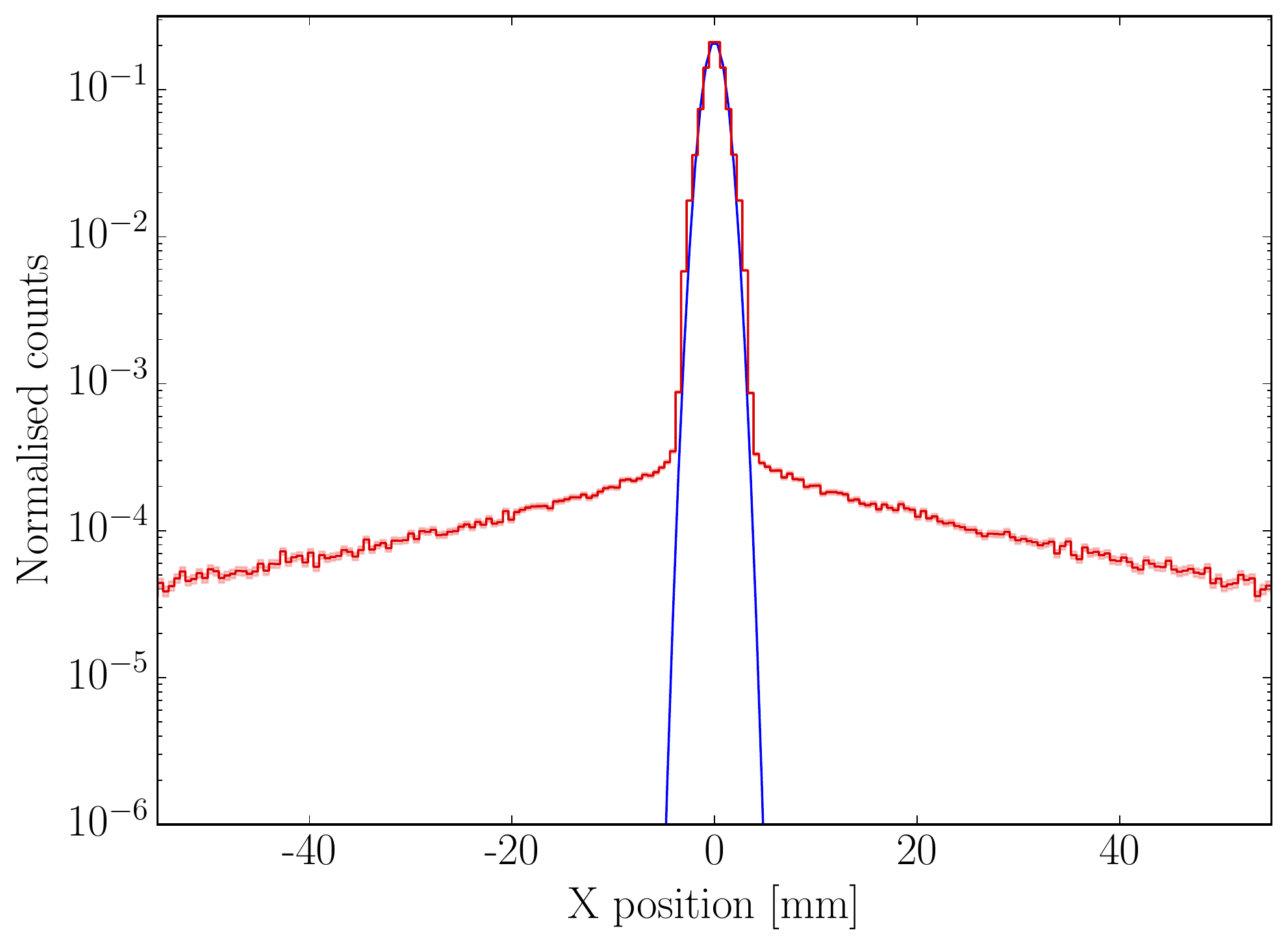}
    \caption{\footnotesize}
    \label{fig:xhit}
  \end{subfigure}%
  \begin{subfigure}{0.5\textwidth}
    \centering
    \includegraphics[width=\textwidth]{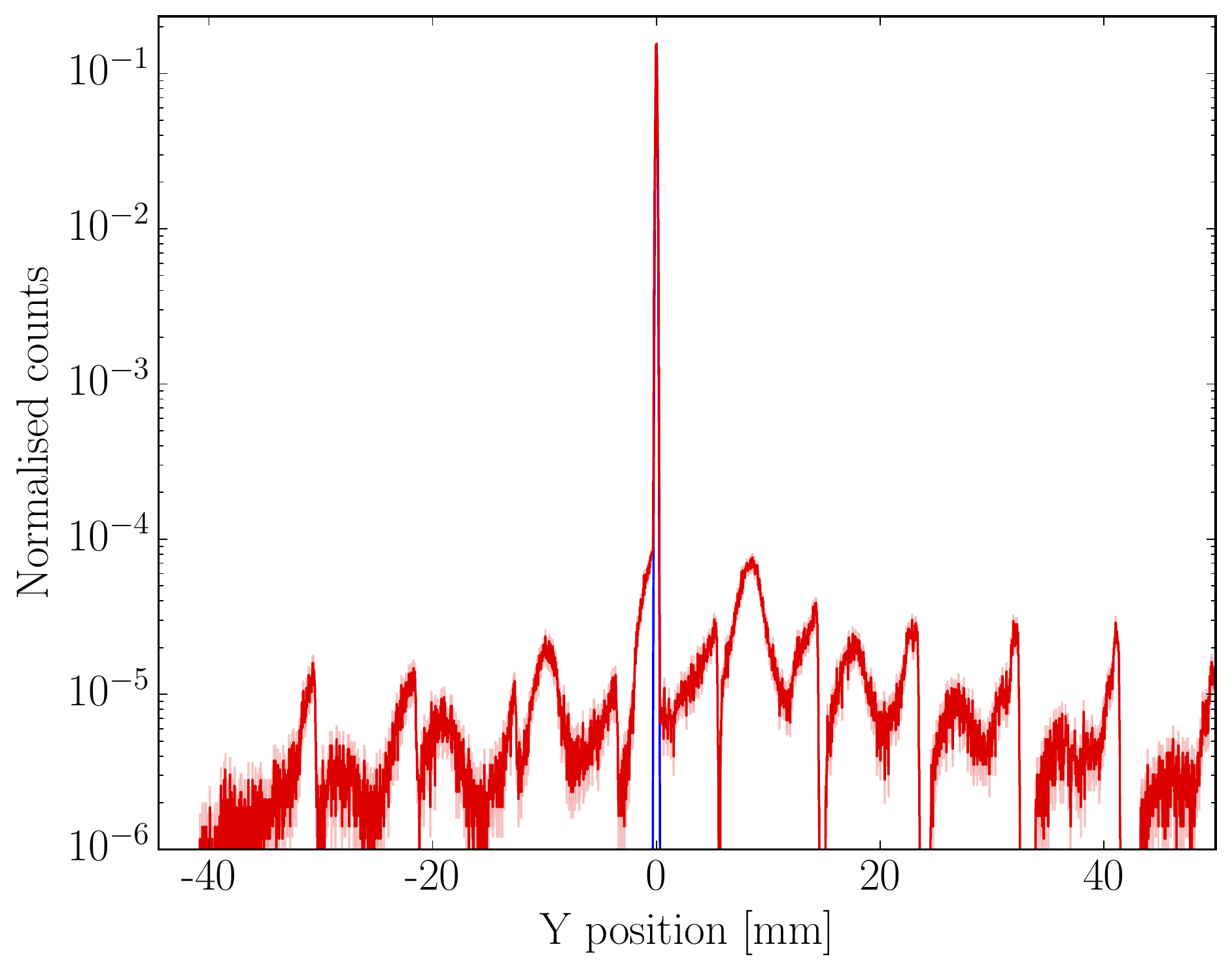}
    \caption{\footnotesize }
    \label{fig:yhit}
  \end{subfigure}
  \caption{\footnotesize  (a) Two-dimensional distribution of hit coordinates,
    projected on the detector window surface. X (b) and Y (c)
    projections of distribution (a) in red. The figures are produced with
    1~\AA~neutron wavelength and a window thickness of
    1~cm. All distributions are
    normalised to 1. A Gauss fit of the central bins (in blue) gives
    $\sigma_X$=0.968~mm (FWHM$^{fit}_X$=2.27~mm) and $\sigma_Y$=0.0667~mm (FWHM$^{fit}_Y$=0.157~mm).} 
  \label{fig:2dhits_xy}
\end{figure}

A visualisation of the scattered hits projected on the detector window appear
in figure~\ref{fig:2dhits_xy}. The projection is necessary as the
window is not vertical but has a 1$^{\circ}$ angle with respect to the
Y-axis. It is also the standard way of experimentally visualising the data

The primary neutrons hit the centre of the distribution in
figure~\ref{fig:2dhits}. The simulation is run for the extreme case
of 1~\AA~for the neutron wavelength and 1~cm for the window thickness,
in order to maximise the scattering effects. The entries away from the
centre make up the scattering
events. The asymmetry of the distribution reflects the asymmetry in
the registration of the detection events due to the orientation of the
blades with respect to the incident beam. The latter effect is better
demonstrable in the projection of
figure~\ref{fig:yhit}. Both figures~\ref{fig:xhit} and \ref{fig:yhit} depict the short-scale
effect attributed to the detector resolution, manifesting itself as the Gaussian smearing
around 0. The extended tails on either side of the distribution represent
the long-scale scattering events. In the case of figure~\ref{fig:yhit} the tails are
``modulated'' by the succession of the blades in the Y-direction, in
addition to the window effects. Such
structures are absent in the X-direction because of the detector
symmetry along the wire length.

A detected neutron is considered misplaced, if the detection occurs
outside the 4$\sigma$ ($\sim$2$\times$FWHM) of the Gaussian fit of the spatial hit
distributions (see figures~\ref{fig:xhit} and~\ref{fig:yhit}), i.e.\,3.87~mm in the direction parallel to the wires
(X-direction) and 0.267~mm in the Y-direction. The $\sigma_X$ and $\sigma_Y$ of
the two fits indicate minimal impact of scattering on resolution
(e.g.\,FWHM$^{detector}_Y$=0.55~mm and FWHM$^{fit}_Y$ $\approx$ 0.3$\times$FWHM$^{detector}_Y$). The fraction of scattered neutrons is estimated by summing up all hits
that fulfil the above condition and dividing
this sum with the total number of detected neutrons, as in the
equation below
\begin{equation}
Fraction = \frac{N_{scattered~neutrons} (|x|\geq 3.87~mm~and~|y|\geq 0.267~mm)}{N_{all~neutrons}}\Bigg\rvert
_{detected}.
\label{fraction_fom}
\end{equation}
This is the figure of merit used in the following
subsections, in contrast to the instrument approach, which evaluates the
peak-to-tail ratio. The current approach yields higher values of
fractional scattering but is important for the detector evaluation.

\FloatBarrier

\subsection{Comparison of simulation with experimental results}

Recently acquired experimental data allow for a comparison with the simulation results. The
measurements~\cite{crisp,royalsoc} were performed at the CRISP neutron reflectometer at the
ISIS neutron and muon source~\cite{isis}. The Geant4 model matches the
geometrical choices of the respective demonstrator, i.e.\,the window
thickness is 2~mm and the converter layer is 4.4~$\mu$m thick in this
particular demonstrator. The neutron
wavelength in the simulation is set to 1~\AA~and the beam is
pencil-like, while the experimental beam profile was approximately
3~mm~$\times$~60~mm. The simulation simplification serves the
purpose of providing a clearer picture of the scattering topology and
is justified by the detector symmetry along the X-axis.

\begin{figure}[!h]
  \centering
  \includegraphics[width=.6\textwidth]{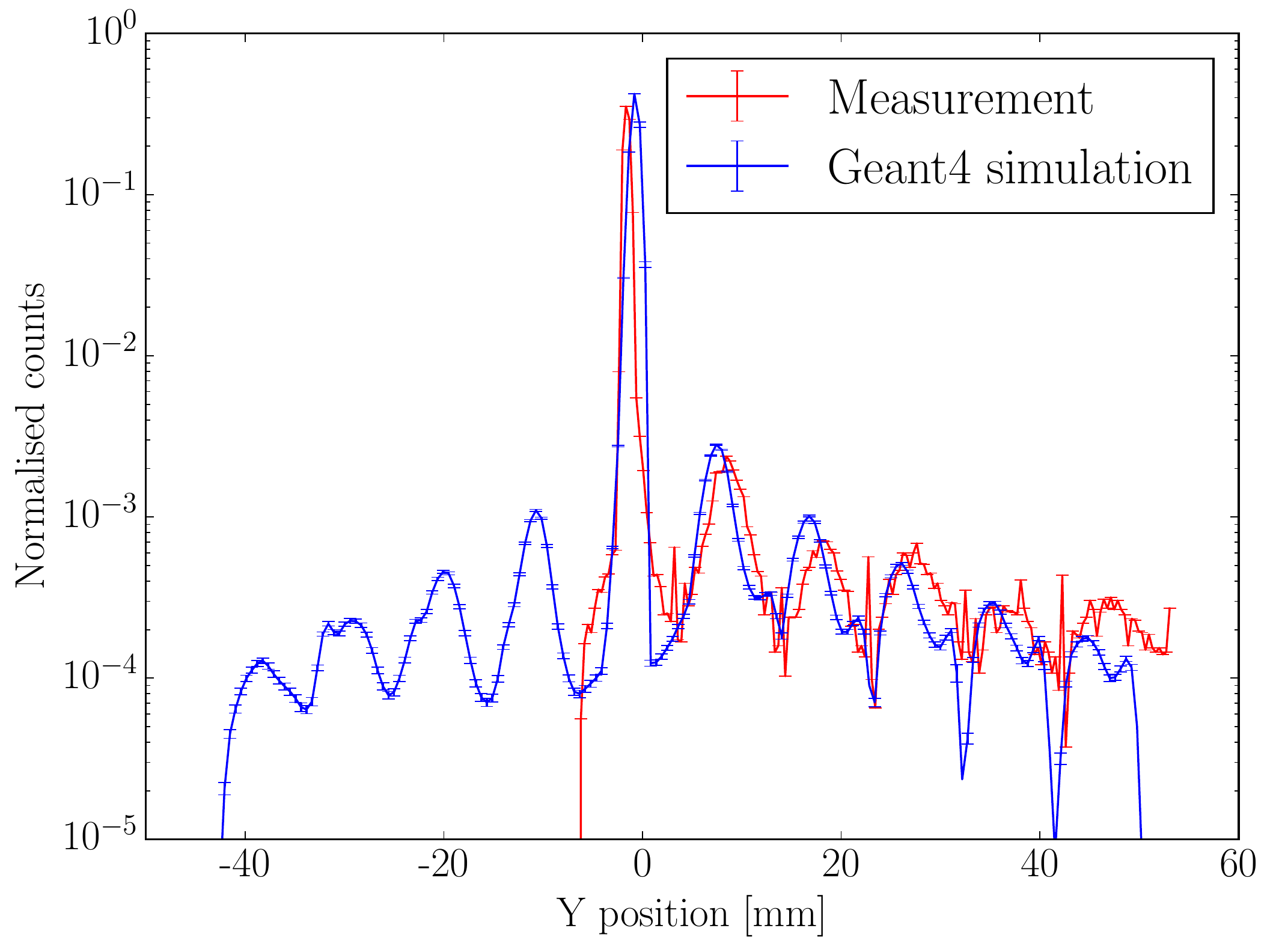}
  \caption{\footnotesize Comparison of Y-position of detected neutrons in measurements
    taken at CRISP~\cite{crisp} and the results of the simulation
    (this work).}   
  \label{fig:crisp} 
\end{figure}

The basis of the comparison is the Y position of the detection events,
as in figure~\ref{fig:yhit} and is shown in figure~\ref{fig:crisp}. The experimental data correspond to a wavelength range of
0.5-2.5~\AA, achieved via the application of a TOF slice~(red distribution of figure~10
in~\cite{crisp}). The simulated coordinates are smeared with a
Gaussian of $\sigma$=0.42~mm (ca.\,2$\times$FWHM)~\cite{crisp,royalsoc}. Only six cassettes were
  experimentally read-out, therefore only six ``peaks'' appear in the
  blue distribution of figure~\ref{fig:crisp}. In addition, three wires from the end of
  each cassette do not contribute to the measurement and have also been accounted for in the simulation by
  excluding the respective Y histogram bins. The distributions are normalised
  to their integral. 

The Geant4 simulation is able of reproducing the shapes, widths and
features of the experimental data. The systematic effects are out of
scope of the current work. The shape agreement indicates that the simulation reproduces
the location of the scattered events and gives confidence that the
topology of scattering is understood in a manner sufficient for the purposes of this study. The fraction of scattered
neutrons in the simulation is 3.5\%. The same fraction estimated from
the experimental data amounts to 3.1\%.

\FloatBarrier

\subsection{Impact of the converter thickness}
\label{sec:boron}

In~\cite{MB2017,crisp} it is argued that an increase of the converter
thickness can respectively increase the fraction of absorbed neutrons,
if set to a value higher than 4.4~$\mu$m (figure~11
of~\cite{crisp}). The motivation is to prevent as many neutrons from reaching the titanium
blade behind the coating, which is the primary contributor to scattering. The detection
efficiency anyway saturates at thicknesses above 3~$\mu$m~\cite{MB2017}. At the same
time, an upper value limit needs to be determined, as it is not
cost-effective, nor good thin-film deposition practice, to arbitrarily increase the boron carbide thickness.

\begin{figure}[!h]   
  \centering
  \begin{subfigure}{0.5\textwidth}
    \centering
    \includegraphics[width=\textwidth]{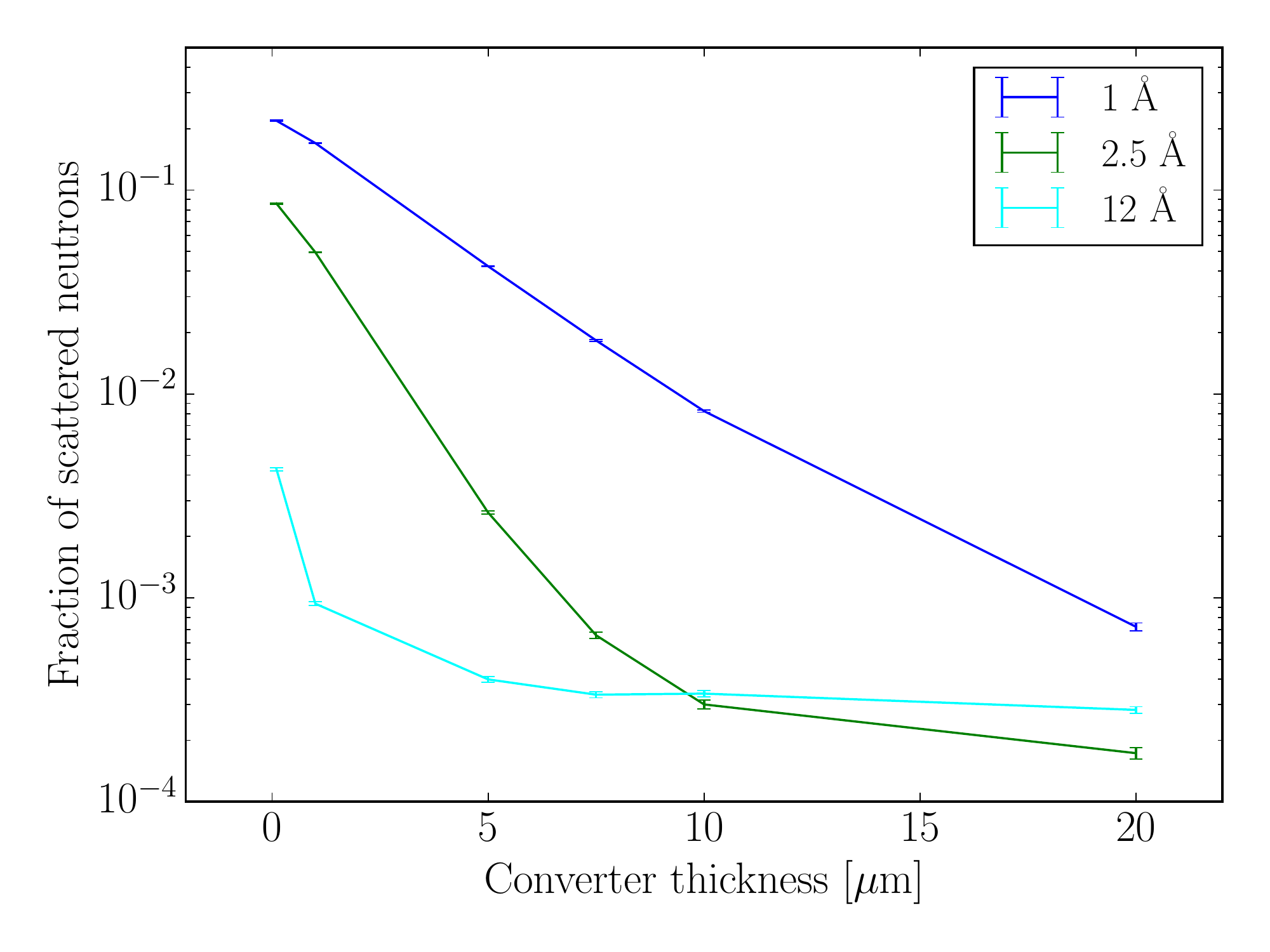}
    \caption{\footnotesize}
    \label{scatteredNeutrons_b10}    
  \end{subfigure}%
  \begin{subfigure}{0.5\textwidth}
    \centering
    \includegraphics[width=\textwidth]{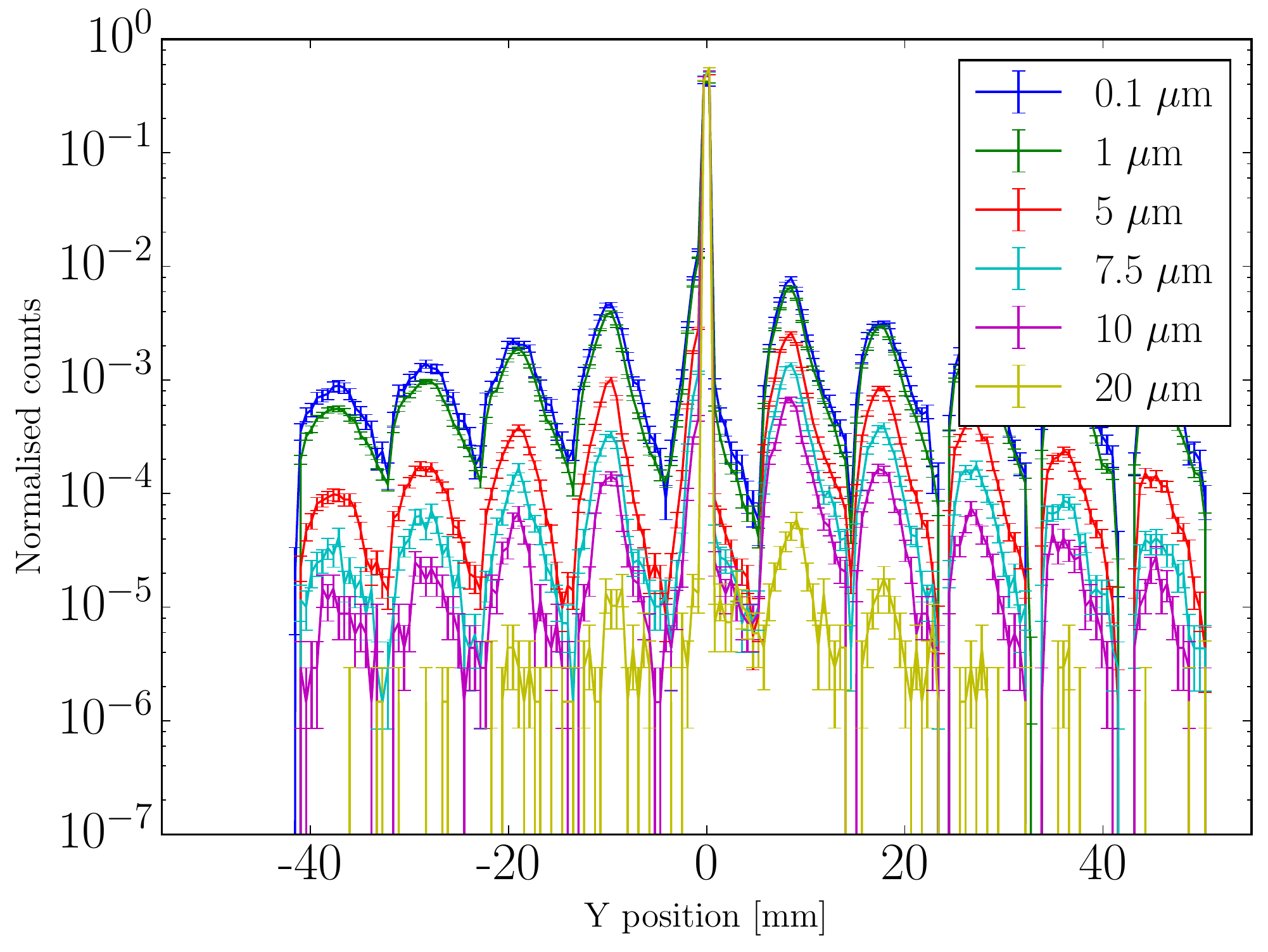}
    \caption{\footnotesize}
    \label{scan_boron}
  \end{subfigure}
  \caption{\footnotesize  (a) Fraction of scattered neutrons as defined in
    equation~\ref{fraction_fom} as a function of converter thickness
    for various neutron wavelengths. (b) Y position of hits for
    various converter thicknesses for a neutron wavelength of
    1~\AA. The entrance window material is set to vacuum. 
  } 
  \label{scan_b10}
\end{figure}

For the study of the converter thickness the material of the detector
window is set to vacuum. Various thicknesses are selected, ranging from
typical to extreme. The results are summarised in table~\ref{table_b10} and
figure~\ref{scatteredNeutrons_b10}, listing the figure of merit
defined with equation~\ref{fraction_fom}.
\begin{table}[!h]
  \caption{\footnotesize Fraction of scattered neutrons in \%, as defined in
    equation~\ref{fraction_fom}, for various converter thicknesses and
    neutron wavelengths.}
  \label{table_b10}
  \centering
  \begin{tabular}{ |c||c|c|c|c|c|c| }
    \hline
    \diagbox[linewidth=0.2pt]{$\lambda$~[\AA]}{c. thickness
      [$\mu$m]} & 0.1 & 1 & 5 & 7.5 & 10 & 20 \\ \hline \hline
    1  & 22.250 & 17.310 & 4.310 & 1.840 & 0.831 & 0.074\\ \hline
    2.5& 8.570  & 5.080  & 0.265 & 0.067 & 0.034 & 0.022\\ \hline
    12 & 0.460  & 0.098  & 0.045 & 0.040 & 0.038 & 0.037\\ \hline
  \end{tabular}
\end{table}

The table and figure values represent the
  scattering which is intrinsic to the detector and is primarily
  attributed to the blade material. Minor contributions come from the
  counting gas, the converter and the wires. Figure~\ref{scan_boron} depicts the
  distribution of the hit Y position for various converter thicknesses.
Clearly, the thicker the coating, the higher
the probability is that all neutrons convert and stop in the first layer -
regardless of being detected, thus minimising scattering effects inside the
detector. Adopting a thickness between 5~$\mu$m and
  10~$\mu$m can reduce the fractional scattering by 1-2 orders of
  magnitude. The trend in figure~\ref{scatteredNeutrons_b10} demonstrates that for
wavelengths above 2.5~\AA~and 4~\AA~which are the lowest limit for FREIA and
ESTIA respectively, a cost-effective choice of converter thickness would be of the order
of 7-8~$\mu$m, for the fraction of scattered neutrons to stay below
10$^{-3}$.

\FloatBarrier

\subsection{Impact of the detector window thickness}

\begin{figure}[!h]
  \centering
  \includegraphics[width=.6\textwidth]{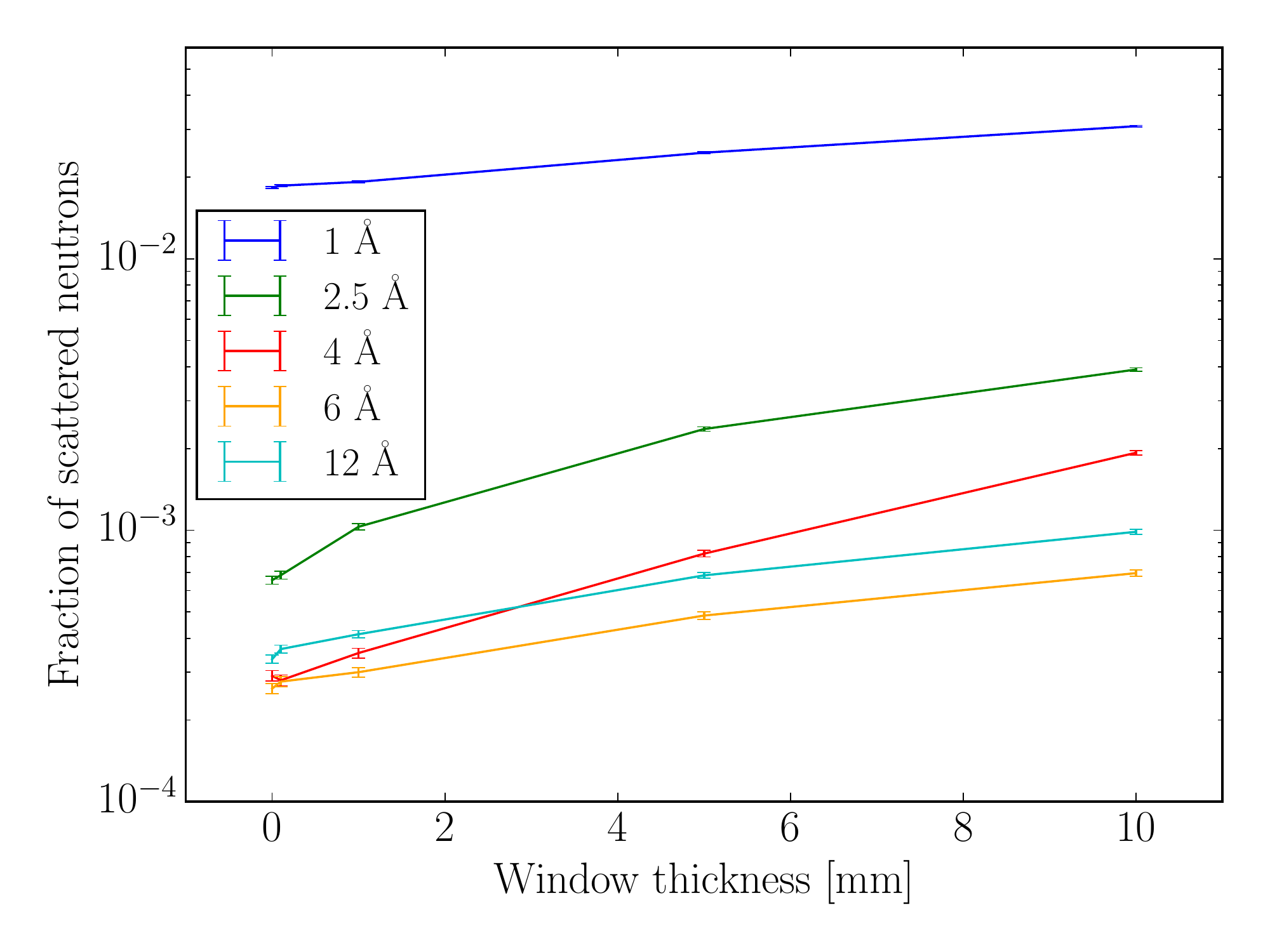}
  \caption{\footnotesize Fraction of scattered neutrons as defined in
    equation~\ref{fraction_fom} as a function of window thickness for
    various wavelengths and a converter thickness of 7.5~$\mu$m.}
  \label{scatteredFractions_wt}
\end{figure}

\begin{figure}[!h]   
  \centering
  \begin{subfigure}{0.5\textwidth}
    \centering
    \includegraphics[width=\textwidth]{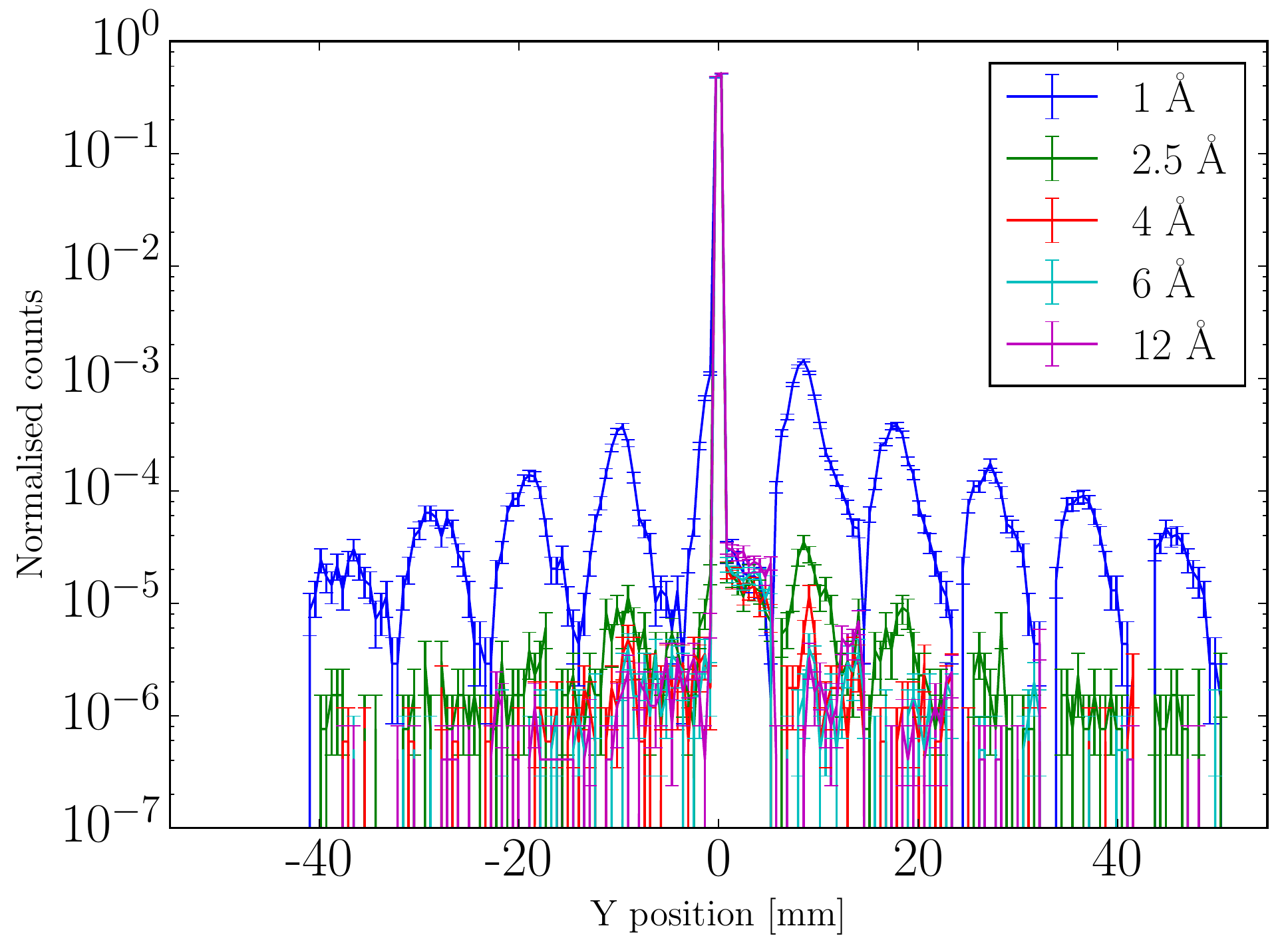}
    \caption{\footnotesize}
    \label{scan_lambda}    
  \end{subfigure}%
  \begin{subfigure}{0.5\textwidth}
    \centering
    \includegraphics[width=\textwidth]{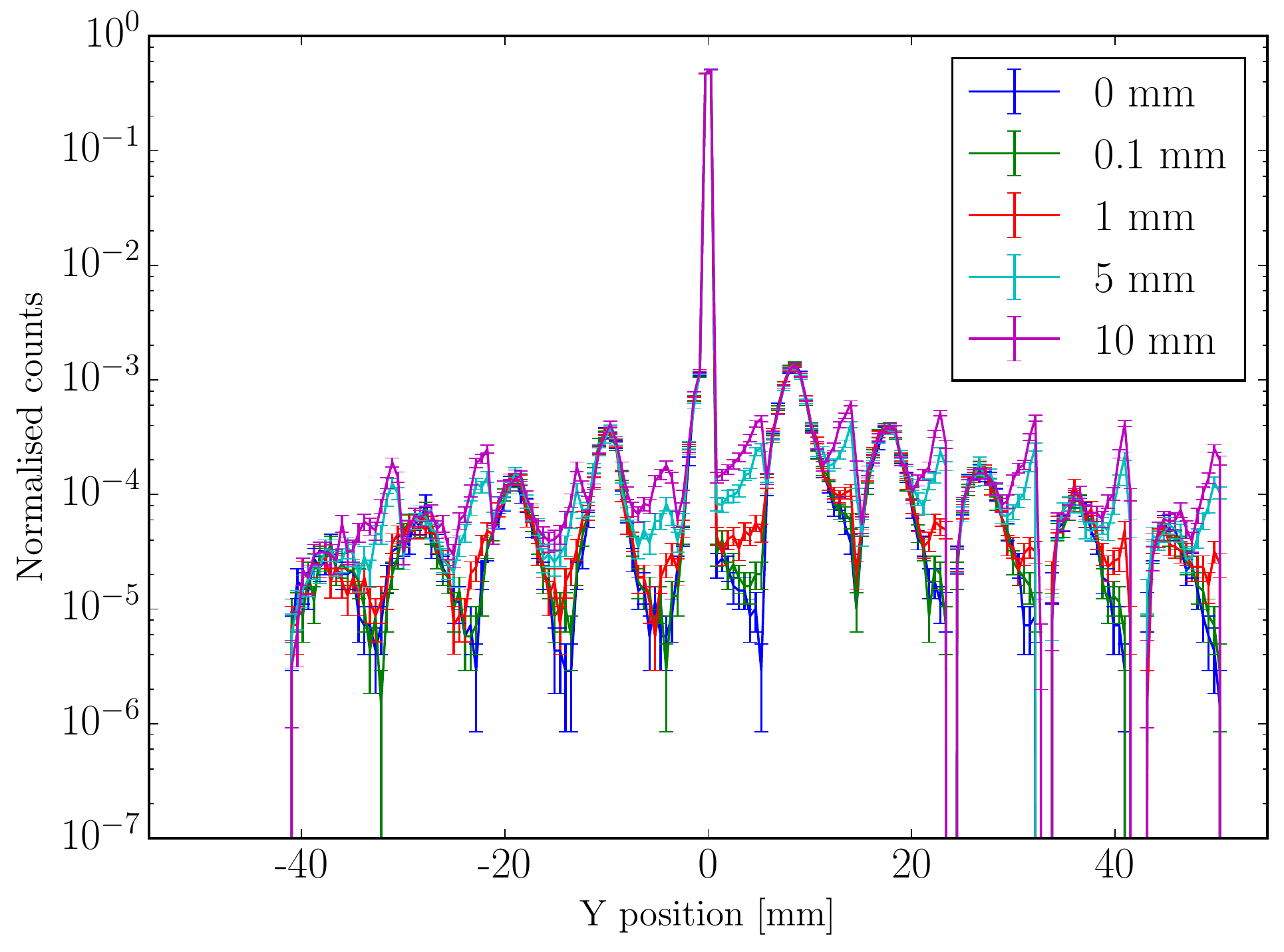}
    \caption{\footnotesize}
    \label{scan_wt}
  \end{subfigure}
  \caption{\footnotesize (a) Y position of hits for various neutron
    wavelengths and no detector window for converter thickness of
    7.5~$\mu$m. (b) Same distribution for 1~\AA~with varying window thicknesses.}
  \label{scan_lambda_wt}
\end{figure}

Following the same methodology, the next parameter to be studied is
the thickness of the Al detector window. This item is an integral
component of the detector, separates the counting gas from the
detector environment and assures its operation in vacuum or atmospheric conditions. Similar studies have been performed with other
detector types~\cite{mg_scattering}. The values selected represent
typical thicknesses used in neutron instruments, in addition to extreme
values. The converter thickness for this series of simulations is now
fixed at 7.5~$\mu$m. The results are summarised in
table~\ref{combinedTable} and figures~\ref{scatteredFractions_wt} and
\ref{scan_lambda_wt}.  Looking at figure~\ref{scatteredFractions_wt},
the scattering effects become more significant for lower neutron
wavelengths. The result is consistent with the trend of the
scattering (see appendix~\ref{app_nc}), in combination with the absorption cross sections of the
materials involved for the specific geometry.   

It is interesting to note that a different, e.g.\,larger
detector geometry, would not only scatter differently but also
register these events differently. Especially for wavelengths below
the Bragg cut-off value of Al and Ti (see
figure~\ref{scat_xsections}), it is possible to imagine a scenario
where the Al window scatters in such a way that although the fraction
of scattered neutrons is higher, the scattering angles are such that
the neutrons are diverted away from the detector active volume. That
is why it is essential to study the impact of the window in
combination with the detector response. Scattering is only an issue if it is detected.

The Multi-Blade detector is intended to operate both in vacuum and a
normal pressure atmosphere. Given the engineering
considerations based on these atmospheric conditions and the fact that
the final detector installed at the ESS reflectometers will have a
window with a size of 500~mm~$\times$~250~mm, engineering studies show that the
window thickness can safely remain below 5~mm for vacuum
operation. Recent developments in the ESTIA design promote a neutron flight
vessel of a 1~bar Ar/$^4$He mixture, in which case the Multi-Blade
detector can be operated with a thin foil (ca.\,25-100~$\mu$m) instead of a window. In summary, for the wavelengths of
interest, the fraction of scattered neutrons at the presence of the
window is within acceptable limits.
\begin{table}[!h]
  \caption{\footnotesize Fraction of scattered neutrons in \%, as defined in
    equation~\ref{fraction_fom}, as a function of window
  thickness and neutron wavelength for a converter thickness of
    7.5~$\mu$m.}
  \label{combinedTable}
  \centering
  \begin{tabular}{ |c||c|c|c|c|c| }
    \hline
    \diagbox[linewidth=0.2pt]{$\lambda$~[\AA]}{window thickness~[mm]} & 0 & 0.1 &1 & 5 & 10 \\ \hline \hline
    1   & 1.840 & 1.870 & 1.970 & 2.480 & 3.060  \\  \hline  
    2.5 & 0.063 & 0.069 & 0.098 & 0.238 & 0.398 \\ \hline 
    4   & 0.030 & 0.030 & 0.039 & 0.088 & 0.187 \\ \hline 
    6   & 0.030 & 0.031 & 0.036 & 0.054 & 0.075 \\ \hline 
    12  & 0.041 & 0.041 & 0.046 & 0.073 & 0.103 \\ \hline
  \end{tabular}
\end{table}


The impact of the detector window is also demonstrable in
figure~\ref{scan_lambda_wt}. In figure~\ref{scan_lambda} the lower peaks on either side of the
centre are at least 4 orders of magnitude below the peak
containing the non-scattered events, especially for the values of interest
above 2.5~\AA. This value satisfies the instrument requirement, the
way it is defined as a peak-to-tail ratio. As expected, the fractional scattering decreases as neutrons get colder. 

\FloatBarrier

\FloatBarrier



\section{Conclusions}

A detailed Geant4 model of the Multi-Blade detector is implemented, in
order to identify and quantify scattering effects. To this end, a parameter scan is
performed focusing mainly on the effects related to the converter
thickness and the detector window. The substrate material and
thickness have been experimentally optimised and therefore the engineering values are used in
the simulation. Various neutron wavelengths of relevance for ESS
reflectometry are used.

A comparison of the simulation with experimental results
obtained at the CRISP reflectometer confirms our understanding of the
detection and scattering topology within the detector. It is shown
that the degradation in spatial resolution due to scattering is smaller than
the detector resolution. The result of
the converter thickness study supports the design choice to increase this
value to 7.5~$\mu$m, with a 2 orders of magnitude gain in terms of
scattering suppression. As for the detector window, its impact is within
the instrument requirements for the values that result from the
engineering studies, in particular for detector operation in vacuum. 

Last but not least, this study selects the amount of fractional
scattering as a figure of merit. It follows a cumulative approach in contrast to the
instrument one, in the sense that the background is integrated over
the entire tail length of the coordinate distributions, while the instrument
requirements are expressed more as a peak-to-tail ratio. This implies
that the cumulative approach presented here leads to an overestimate
of the scattering effects by at least half an order of magnitude. For
neutron wavelengths above 2.5~\AA, the peak-to-tail ratio is higher
than 4 orders of magnitude, which means that the current
implementation of the MultiBlade detector more than satisfies the ESS
reflectometry needs.


\acknowledgments

The authors acknowledge the support from the EU Horizon2020
BrightnESS grant 676548~\cite{brightness}. This work was supported by
the \'UNKP-17-2 new national excellence program of the Hungarian
Ministry of Human Capacities. The authors would like to thank ISIS,
Didcot, UK and
BNC, Budapest, Hungary for the beam time they made available. Computing resources were provided by the DMSC
Computing Centre~\cite{dmsc}. G\'abor Galg\'oczi would like to thank
Bal\'azs \'Ujv\'ari for the fruitful discussions they had.




\appendix
\section{NCrystal cross sections}
\label{app_nc}
For the correct treatment of the interaction of thermal and
cold neutrons with several single crystals, poly-crystalline materials
and powders, the NCrystal library~\cite{ncrystal,icns} is used in this
work. The treatment includes both coherent elastic (Bragg) diffraction and various models for the inelastic
scattering. The library is publicly available for use under
a highly liberal open source license (Apache 2.0) and already
interfaces with several Monte Carlo packages, e.g. McStas~\cite{mcstas1,mcstas2}, ANTS2~\cite{ants2} and Geant4. Its
flexible interface though allows for easy integration with other packages.

\begin{figure}[!b]   
  \centering
  \begin{subfigure}{0.5\textwidth}
    \centering
    \includegraphics[width=\textwidth]{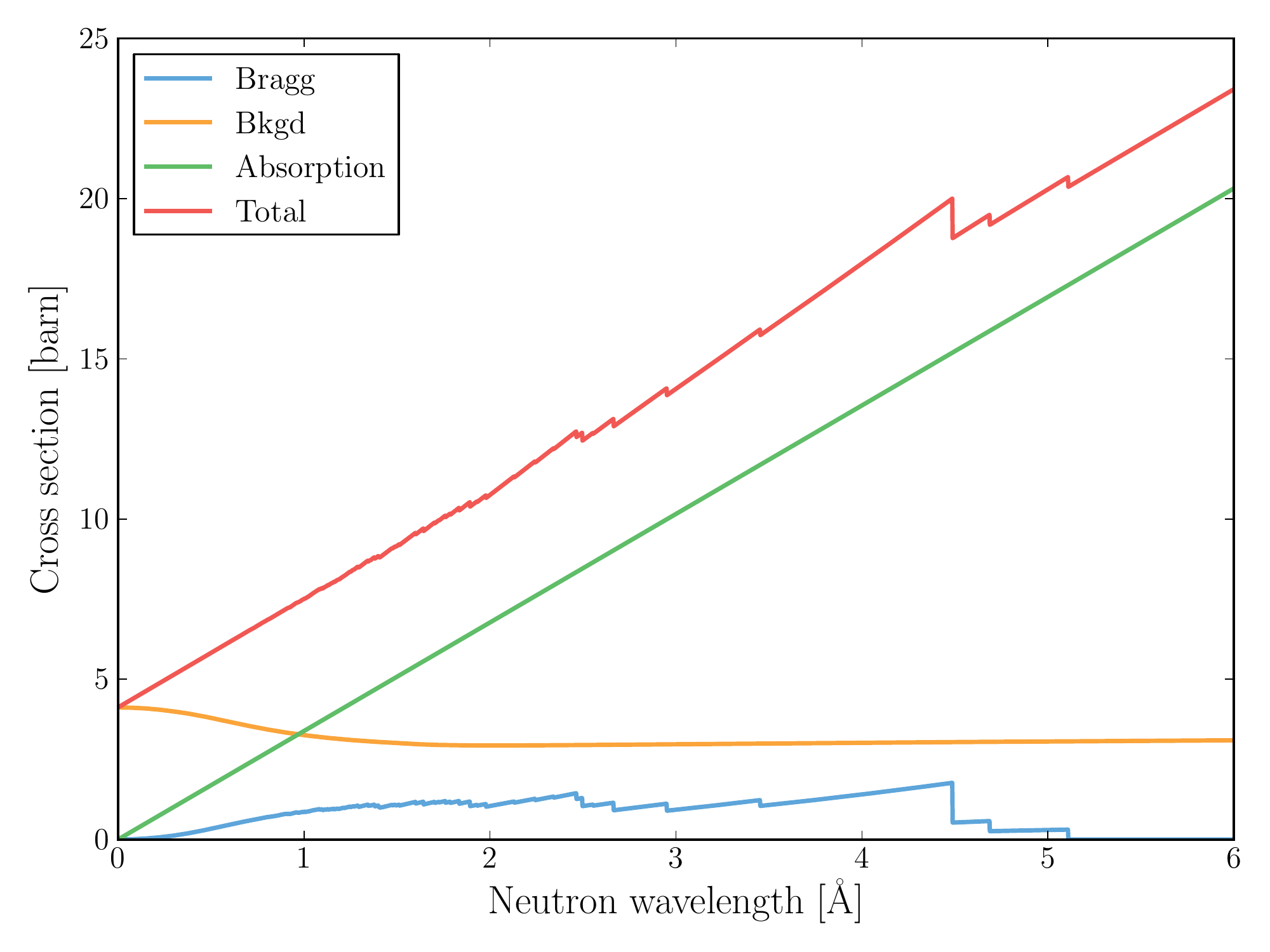}
    \caption{\footnotesize}
    \label{ti_ncrystal}    
  \end{subfigure}%
  \begin{subfigure}{0.5\textwidth}
    \centering
    \includegraphics[width=\textwidth]{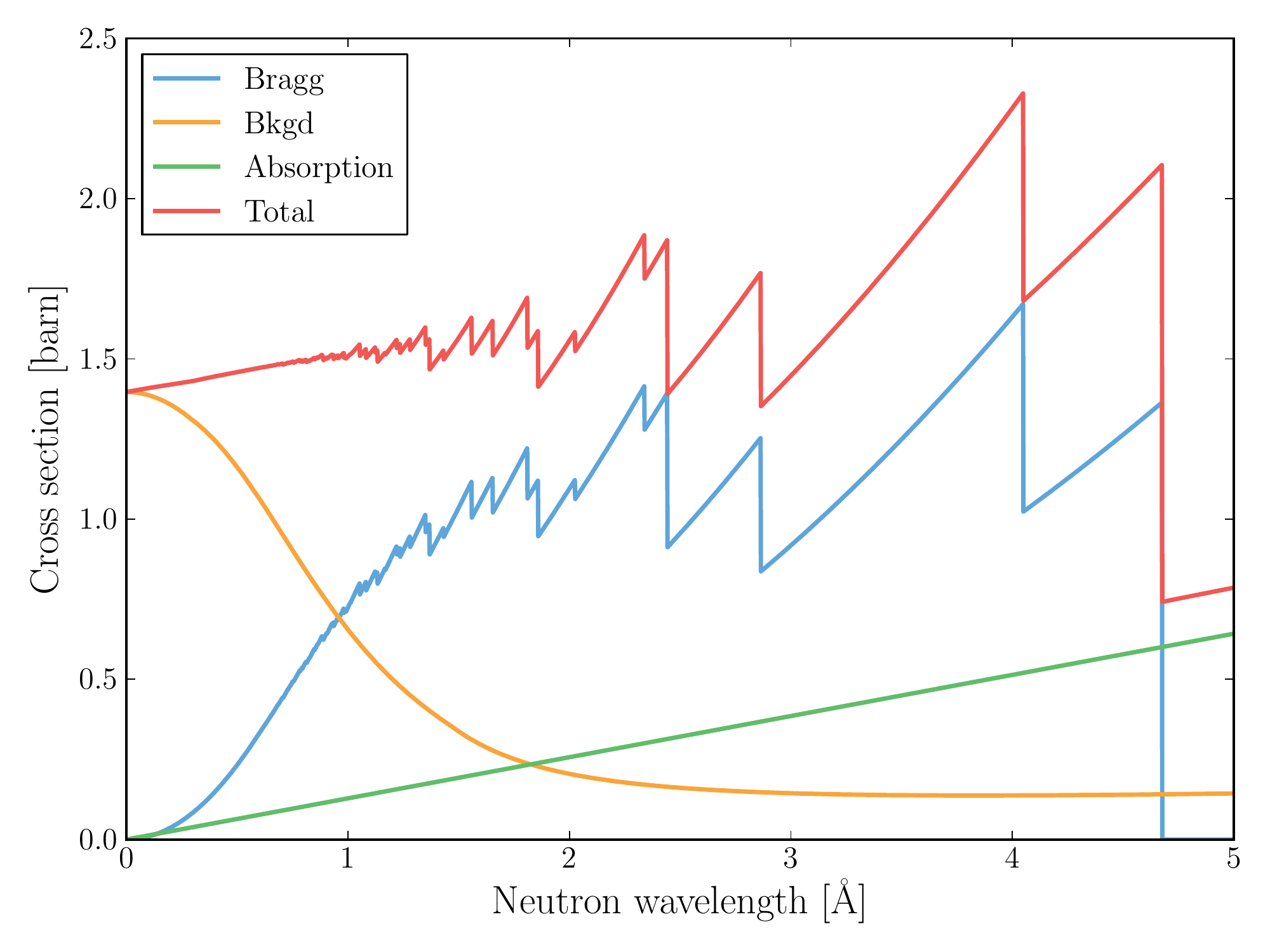}
    \caption{\footnotesize}
    \label{al_ncrystal}
  \end{subfigure}
  \caption{\footnotesize (a) Total scattering and absorption cross
    sections of Ti and (b) Al vs.\,neutron wavelength. ``Bkgd''
    refers to coherent inelastic, elastic incoherent and
    inelastic incoherent processes.}
  \label{scat_xsections}
\end{figure}

Two application examples that are used in this work are Ti and Al, presented
in figure~\ref{scat_xsections}. The reproduction of the respective
scattering cross sections would not be possible without NCrystal. The
library has been validated against available experimental data. A collection of existing
crystalline structures supported by NCrystal can be found in~\cite{data_library}.

\begin{figure}[!h]
  \centering
  \begin{subfigure}{0.5\textwidth}
    \centering
    \includegraphics[width=\textwidth]{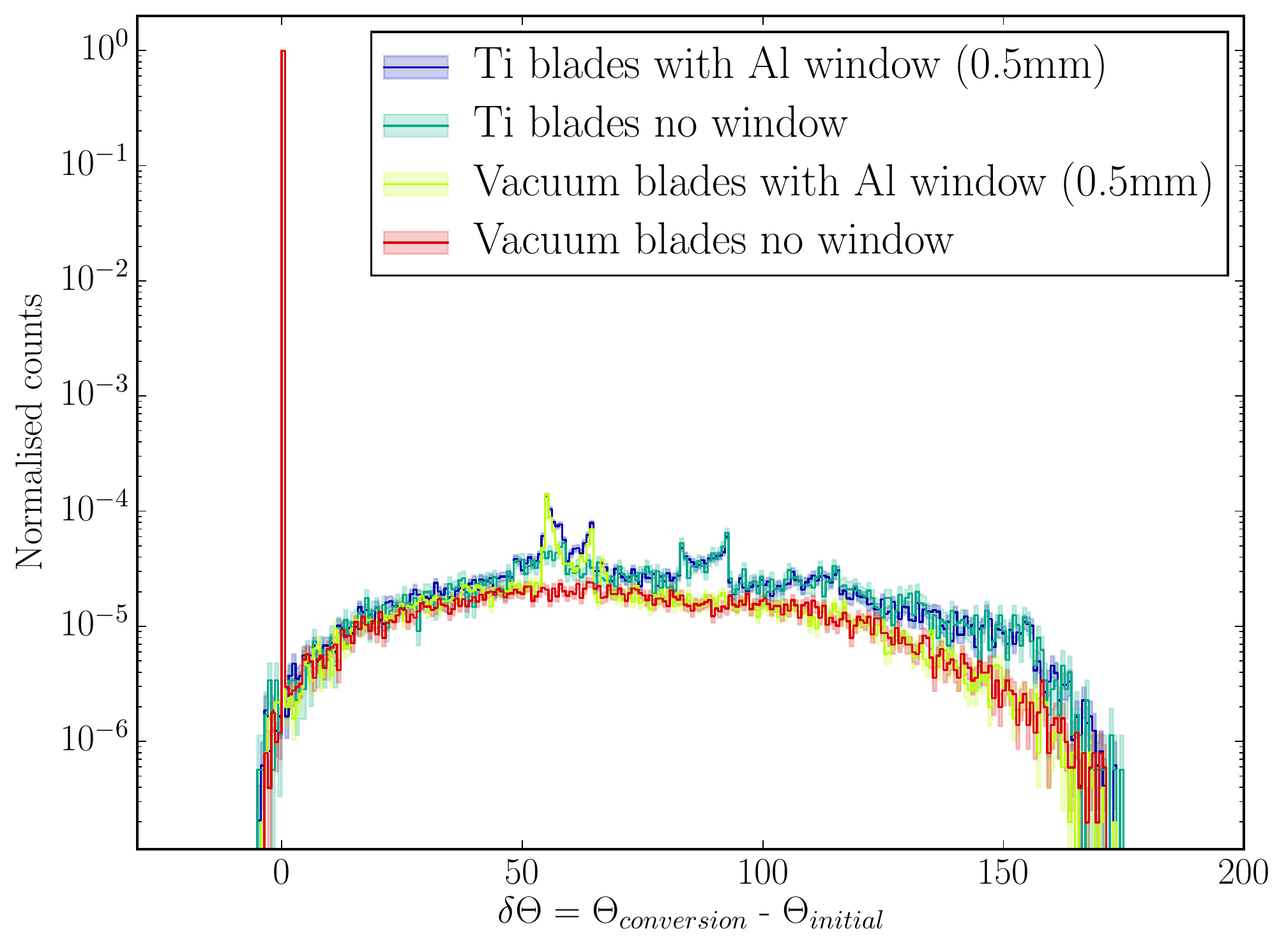}
    \caption{\footnotesize}
    \label{dtheta}
  \end{subfigure}%
  \begin{subfigure}{0.475\textwidth}
    \centering
    \includegraphics[width=\textwidth]{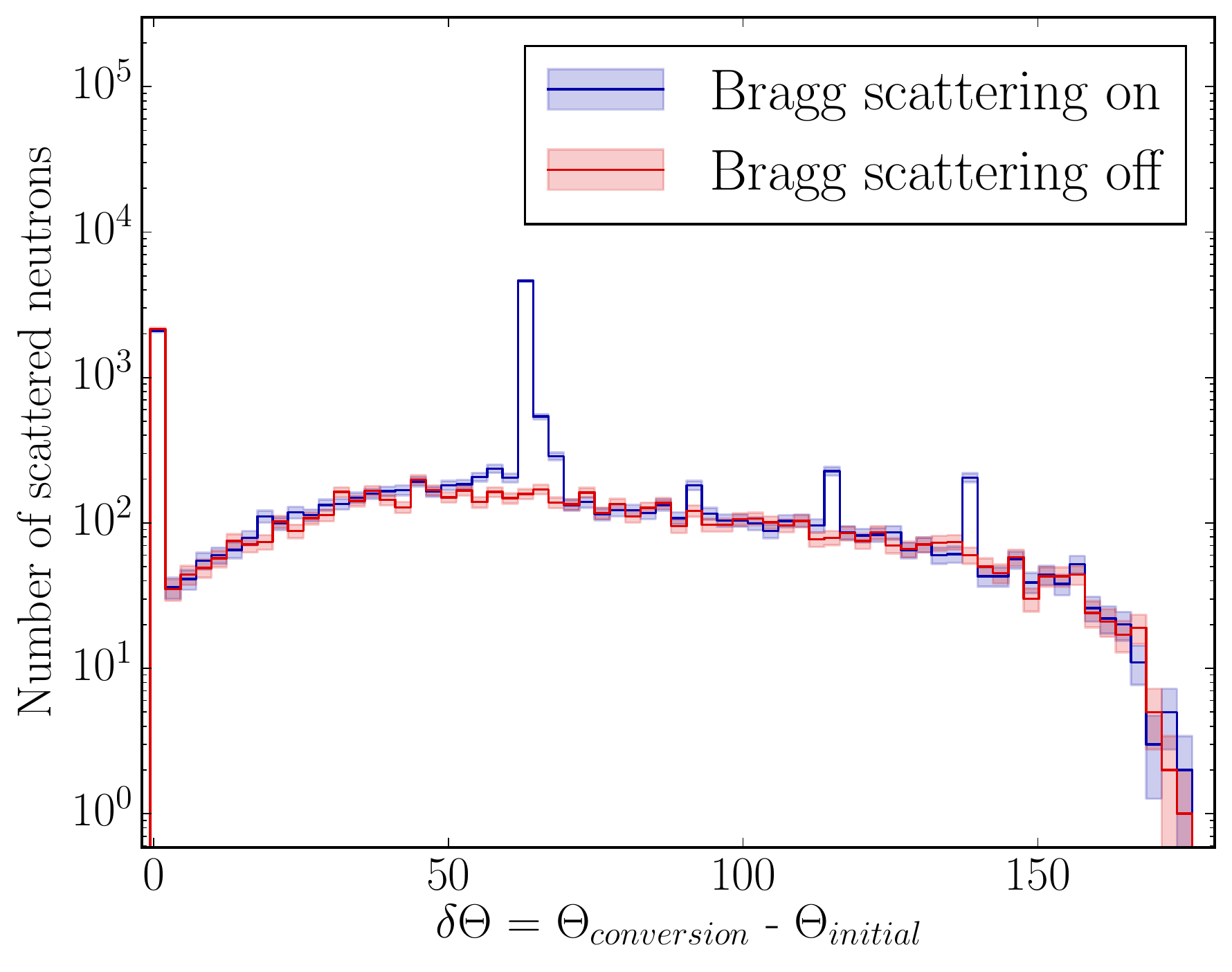}
    \caption{\footnotesize }
    \label{bragg_nonbragg}
  \end{subfigure}
  \caption{(a) Difference between the
      conversion polar angle and the primary polar angle for neutrons
      for a neutron wavelength of 2.5~\AA. Coherent elastic scattering on the
      Al detector window and the Ti blades is responsible for the
      structures appearing around 50$^{\circ}$, 80$^{\circ}$ and
      150$^{\circ}$. (b) Same distribution for ``background'' neutrons
      only and with both Al and Ti present. The red distribution
      disables Bragg scattering in the detector materials.}
\end{figure}

A visual demonstration of the various types of scattering in
the detector blade and window appears in figure~\ref{dtheta}. Coherent
elastic scattering takes place when the neutron wavelength is below the Bragg
cut-off value of the materials it is transported in (4.67~\AA~for Al
and 5.14~\AA~for Ti). What the figure presents for a neutron wavelength of
2.5~\AA~is the difference between the polar angle of a neutron at its
conversion point - calculated from its momentum vectors - and the initial polar angle with which it is
generated:
\begin{equation}
\delta\Theta = \Theta_{conversion}-\Theta_{initial}.
\end{equation}
The red distribution contains only the contributions from
gas and converter, as the blade and window materials are set to
vacuum. No Bragg scattering takes place with these conditions, as
the remaining materials are not crystalline. Once the blade material is set to Ti, structures from the Debye-Scherrer
cones become apparent (in
green). Similarly, the Bragg scattering from the Al window appears in
the remaining two distributions (yellow, blue).

By selecting ``background'' neutrons, as defined in the denominator of
equation~\ref{fraction_fom}, and switching off Bragg scattering in the simulation, it is estimated that about 30\% of the
neutrons scatter coherently elastically for 2.5~\AA~and a 2~mm Al window (see figure~\ref{bragg_nonbragg}).

\end{document}